\documentclass[draft]{agujournal2019}
\usepackage{url} 
\usepackage{lineno}
\usepackage[inline]{trackchanges} 
\usepackage{soul}
\usepackage{booktabs}
\usepackage{multirow}

\draftfalse

\journalname{JGR: Space Physics}

\begin{document}

%
%


\title{Influence of the Jovian current sheet models on the mapping of the UV auroral footprints of Io, Europa, and Ganymede}

%
%




\authors{J. Rabia\affil{1}, Q. Nénon\affil{1}, N. André\affil{1}, V. Hue\affil{2}, D. Santos-Costa\affil{3}, A. Kamran\affil{1} and M. Blanc\affil{1,2}}


\affiliation{1}{Institut de Recherche en Astrophysique et Planétologie, CNRS-UPS-CNES, Toulouse, France}
\affiliation{2}{Aix-Marseille Université, CNRS-CNES, Institut Origines, LAM, Marseille, France}
\affiliation{3}{Southwest Research Institute, San Antonio, TX, USA}





\correspondingauthor{J. Rabia}{jonas.rabia@irap.omp.eu}




\begin{keypoints}
\item We compare the ability of two current sheet models, coupled to JRM33, to reproduce a wide diversity of Galileo and Juno observations
\item The CON2020 model better matches the observations in the inner magnetosphere, including magnetic field measurements and UV auroral footprints 
\item In the outer magnetosphere, the consideration of local time effects by the KK2005 model provides a better agreement against observations
\end{keypoints}

%
%

%
%


\begin{abstract}

The in-situ characterization of moon-magnetosphere interactions at Jupiter and the mapping of moon auroral footpaths require accurate global models of the magnetospheric magnetic field. In this study, we compare the ability of two widely-used current sheet models, Khurana-2005 (KK2005)  and Connerney-2020 (CON2020) combined with the most recent internal magnetic field model of Jupiter (JRM33) to match representative Galileo and Juno measurements acquired at low, medium, and high latitudes. With the adjustments of the KK2005 model to JRM33, we show that in the outer and middle magnetosphere (R$>$15$\textit{R}_{\mathrm{J}}$), JRM33+KK2005 is found to be the best model to reproduce the magnetic field observations of Galileo and Juno as it accounts for local time effects. \newline JRM33+CON2020 gives the most accurate representation of the inner magnetosphere. This finding is drawn from comparisons with Juno in-situ magnetic field measurements and confirmed by contrasting the timing of the crossings of the Io, Europa, and Ganymede flux tubes identified in the Juno particles data with the two model estimates. JRM33+CON2020 also maps more accurately the UV auroral footpath of Io, Europa, and Ganymede observed by Juno than JRM33+KK2005. The JRM33+KK2005 model predicts a local time asymmetry in position of the moons' footprints, which is however not detected in Juno's UV measurements. This could indicate that local time effects on the magnetic field are marginal at the orbital locations of Io, Europa, and Ganymede. Finally, the accuracy of the models and their predictions as a function of hemisphere, local time, and longitude is explored. 
\end{abstract}

%
%

%


%
%
%
%

\section{Introduction}

The four Galilean moons orbit deep inside the Jovian magnetosphere where the closed magnetic field line topology electrodynamically couples the moons to the atmosphere of the gas giant. In particular, the quasi co-rotating Jovian plasma constantly overtakes the moon's orbital motion. This interaction gives rise to a variety of physical processes, including the generation of Alfvén waves. The Alfvén waves can propagate along the magnetic field lines and accelerate charged particles. One of the manifestations of such interactions is the parallel acceleration of electrons that precipitate in the atmospheric loss cone and trigger peculiar auroral UltraViolet (UV) emissions. As shown in \citeA{hue_2023} and references therein, the brightest of these emissions resembles a spot and likely results from electrons accelerated in the Main Alfvén wing (MAW) of the moon-magnetosphere interaction. Another spot can sometimes be observed upstream of the MAW spot and likely corresponds to the TransHemispheric Beam (TEB), which is the reflected counterpart of the opposite hemisphere MAW \cite{bonfond_2008}. Downtail of the MAW, an extended UV auroral tail is observed and may reveal electron acceleration associated with reflected Alfvén waves (RAW) \cite{bonfond_2017,bonfond_2017b,mura_2018,moirano_2021}. In this framework, an accurate magnetic field model is a key to identify when Juno crosses the TEB, MAW, or RAW of a moon, and thus when Juno can observe in situ the accelerated charged particles and the underlying acceleration processes at work. 

The powerful dynamo of Jupiter generates an "internal" magnetic field \cite<e.g.>{moore2018complex}. The magnetic field measurements from eight of the spacecraft that explored in situ the magnetosphere of Jupiter (Pioneer 10-11, Voyager 1-2, Ulysses, Galileo, New Horizons, and Juno) have been previously employed to model the morphology and intensity of the Jovian internal magnetic field. Initial models like the O6 \cite{connerney_1992}, VIP, VIT4 \cite{connerney_1998}, and VIPAL \cite{hess_2011} also attempted to constrain the higher-degree field of the planet by using the position of the Io UV and infrared auroral footprints \cite{atreya_1977}. Later, the ISaAC model \cite{hess_2017} brought additional constraints by using the position of the UV auroral footprints of Europa and Ganymede \cite{bonfond_2012}. The most recent models of the internal magnetic field model, JRM09 and JRM33 \cite{connerney_2018,connerney_2022}, include the closest ever magnetic field measurements obtained by Juno during its first 9 and 33 orbits around the planet.

In addition to the internal magnetic field, Pioneer 10 was the first spacecraft to reveal the presence of a plasma disk confined near the magnetic equator \cite{smith_1974}, where electric currents flow and induce an external magnetic field that adds to Jupiter’s intrinsic magnetic field. The external magnetic field greatly distorts the shape of the magnetic field lines at the orbital distances of the Galilean moons. This contribution to the total magnetic field strength becomes increasingly significant as we move from the orbit of Io ($B_{\mathrm{Ext}}/{B_{\mathrm{Int}}}$ $\approx$ 0.06) to that of Europa ($B_{\mathrm{Ext}}/{B_{\mathrm{Int}}}$ $\approx$ 0.2) and Ganymede ($B_{\mathrm{Ext}}/{B_{\mathrm{Int}}}$ $\approx$ 0.4). The combination of a current sheet model and an intrinsic magnetic field model is therefore critical in order to model the motion of charged particles and the path of electromagnetic waves in Jupiter’s magnetosphere. In the rest of the paper, we refer to the term "current sheet" only for the purpose of investigating the external magnetic fields induced by the latter. Models using the plasma populations and their distributions along the magnetic equatorial plane and their temporal variations as a function of forces in Jupiter's magnetosphere to compute the current systems and the induced magnetic field \cite{caudal_1986, achilleos_2018, millas_2023} will not be considered in this study. We only use empirical models that reproduce the mean magnetic field in the jovian magnetosphere.

In the present paper, we compare two current sheet models widely used by the scientific community in recent years. The first one was developed following the Galileo mission and is described in \citeA{khurana_2005}. The second model considered here is presented in \citeA{connerney_2022} who updated the \citeA{connerney_1981} model with Juno measurements. In Section 2, we introduce the two models and their characteristics. We then compare the modeling results obtained with the two models against in situ magnetic field measurements obtained by the Galileo and Juno spacecraft in Section 3. A comparison of the models with remote observations of the auroral footprints of the Galilean moons obtained by Juno \cite{hue_2023} is presented in Section 4. In Section 5, we contrast the time estimates of the moons' flux tube crossings derived from the two models with in situ observations from Juno. Finally, the main conclusions and implications for future studies are summarized in Section 6.  

\section{Model characteristics and magnetic field line mapping method}

For the sake of comparison we combine the two current sheet models with the same JRM33 reference model for the internal magnetic field of Jupiter. JRM33 is based on Juno measurements obtained during the first 33 orbits of the mission \cite{connerney_2022}. The 13th-order coefficients of the model are used in the following calculations. 

The current sheet model developed by \citeA{connerney_2020}, with Juno data obtained within 30 $\textit{R}_{\mathrm{J}}$ of the planet (1 $\textit{R}_{\mathrm{J}}$ = 71,492 km), is a successor to the CAN81 model \cite{connerney_1981}, which used measurements from the Voyager 1, Voyager 2 and Pioneer 10 probes.  Hereafter referred to as CON2020, it consists of an axisymmetric disk with inner and outer radii of 7.8 $\textit{R}_{\mathrm{J}}$ and 51.4 $\textit{R}_{\mathrm{J}}$, respectively. This disc has a constant half thickness of 3.6 $\textit{R}_{\mathrm{J}}$ and is tilted by 9.3$^{\circ}$ with respect to Jupiter’s rotation axis towards $\lambda_{\mathrm{S3RH}}$ = 155.8$^{\circ}$. As the longitudinal symmetry assumption is reasonable in the inner and middle magnetosphere ($<$ 30  $\textit{R}_{\mathrm{J}}$) \cite{khurana_1992}, the use of this model to magnetically map the footprints of the Galilean moons is relevant. The model is built around 5 parameters that define the shape and the orientation of the current sheet, whose values are given above, and two parameters, the magnetodisc current constant $\mu_{0}I/2$ and the radial current parameter $\mu_{0}I_{rad}/2\pi$, which control the current intensity within the current sheet. These can be adjusted for each spacecraft orbit to take into account the temporal variability of the current intensity and associated magnetic field. Specifically, \citeA{connerney_2020} provide in their Table 2 the current parameters for each of the first 24 orbits of Juno independently. In this study, we use the values that best fit the observations of these perijoves together, i.e., $\mu_{0}I/2$ = 139.6 nT and $\mu_{0}I_{rad}/2\pi$ = 16.7 nT. Note that all the calculations presented in this study have also been performed using the minimum and maximum values of the parameters $\mu_{0}I/2$ and $\mu_{0}I_{rad}/2\pi$ of the CON2020 model \cite<see Table 2 in>{connerney_2020} and parameters derived by \citeA{vogt_2017} based on Galileo magnetic field measurements in order to account for the short- and long-term variability of the current sheet properties, respectively. This did not lead to significant changes in the results displayed in the following sections. The CON2020 model is entirely independent of the considered intrinsic magnetic field model and can therefore be coupled to any internal field model without any modifications.

The second model studied, presented in \citeA{khurana_1997} and \citeA{khurana_2005}, hereafter referred to as KK2005, was derived from magnetic field measurements of Pioneer 10 and 11, Voyager 1 and 2, Ulysses, and Galileo missions obtained between 5 $\textit{R}_{\mathrm{J}}$ and 100 $\textit{R}_{\mathrm{J}}$. This model intends to reproduce the local time asymmetry of the magnetic field in the outer magnetosphere \cite{khurana_1992,khurana_2022}. The structure of the current sheet is described in \citeA{khurana_2005}. This study used Galileo data to improve the current sheet structure presented in \citeA{khurana_1992}, derived from Voyager 1,2 and Pioneer 10 data. The magnetic field induced by the electric currents flowing in the current sheet is reproduced thanks to an Euler potential model described in \citeA{khurana_1997}. Based on a rigid structure inside the middle magnetosphere, the model takes into account the Sun’s position, i.e., the local time (LT), to progressively bend the current sheet which becomes parallel to the solar wind direction at large radial distances downtail of Jupiter. Thanks to this adjustment, this model intends to remain valid even in the outermost regions of the magnetosphere. This current sheet was originally developed to be coupled to the VIP4 magnetic field model. However, we will show that it can be made compatible with the JRM33 model by updating the Gauss coefficients $g_{1}^{0}$, $g_{1}^{1}$, and $h_{1}^{1}$, which govern the direction and magnitude of the dipolar contribution, with those given in \citeA{connerney_2022}.  The update of these coefficients gives to the current sheet a new inclination and direction that are identical with those of the JRM33 dipole axis. The calculation of the total magnetic field is then divided into two parts, the first being the calculation of the magnetic field induced by the current sheet and the dipole part of Jupiter's intrinsic magnetic field. Finally, the second calculation is the non-dipolar contribution, i.e. that generated by the higher-degree terms, is derived using the standard JRM33 model, with coefficients $g_{1}^{0}$, $g_{1}^{1}$ and $h_{1}^{1}$, set to zero to avoid duplicating the calculation of the dipolar part. This procedure enables KK2005 to be coupled with JRM33 or any other internal magnetic field model,  making it possible to use a higher-degree model, thus more closely reproducing Jupiter's intrinsic magnetic field. 

We provide a Python version of the JRM33+KK2005 model we implemented and used in this study in the Supplementary Information file. We converted the initial KK2005+VIP4 model available in IDL to Python language before implementing the JRM33+KK2005 model. Figures S1 and S2 compare the results obtained with the IDL and Python models, ensuring the correct transcription and operation of the Python KK2005+VIP4 model. We also provide in Figures S3 and Figure S4 a comparison between the JRM33+KK2005 and KK2005+VIP4 model ; the latter will not be used in the rest of the study.   

We note that the magnetic field generated by the Chapman-Ferraro currents flowing through the magnetopause is taken into account in the KK2005 model, but not in the CON2020 model. However, since this contribution to the total magnetic field strength is negligible in the inner and middle magnetosphere \cite<B$_{\theta_{\mathrm{MP}}}$ $<$ 1 nT ;>{saur_2018}, we do not couple the CON2020 model with a magnetopause model.

\section{Comparison of the two current sheet models against in situ Galileo and Juno measurements}

\begin{figure}[h!]
    \centering
    \includegraphics[width=\textwidth]{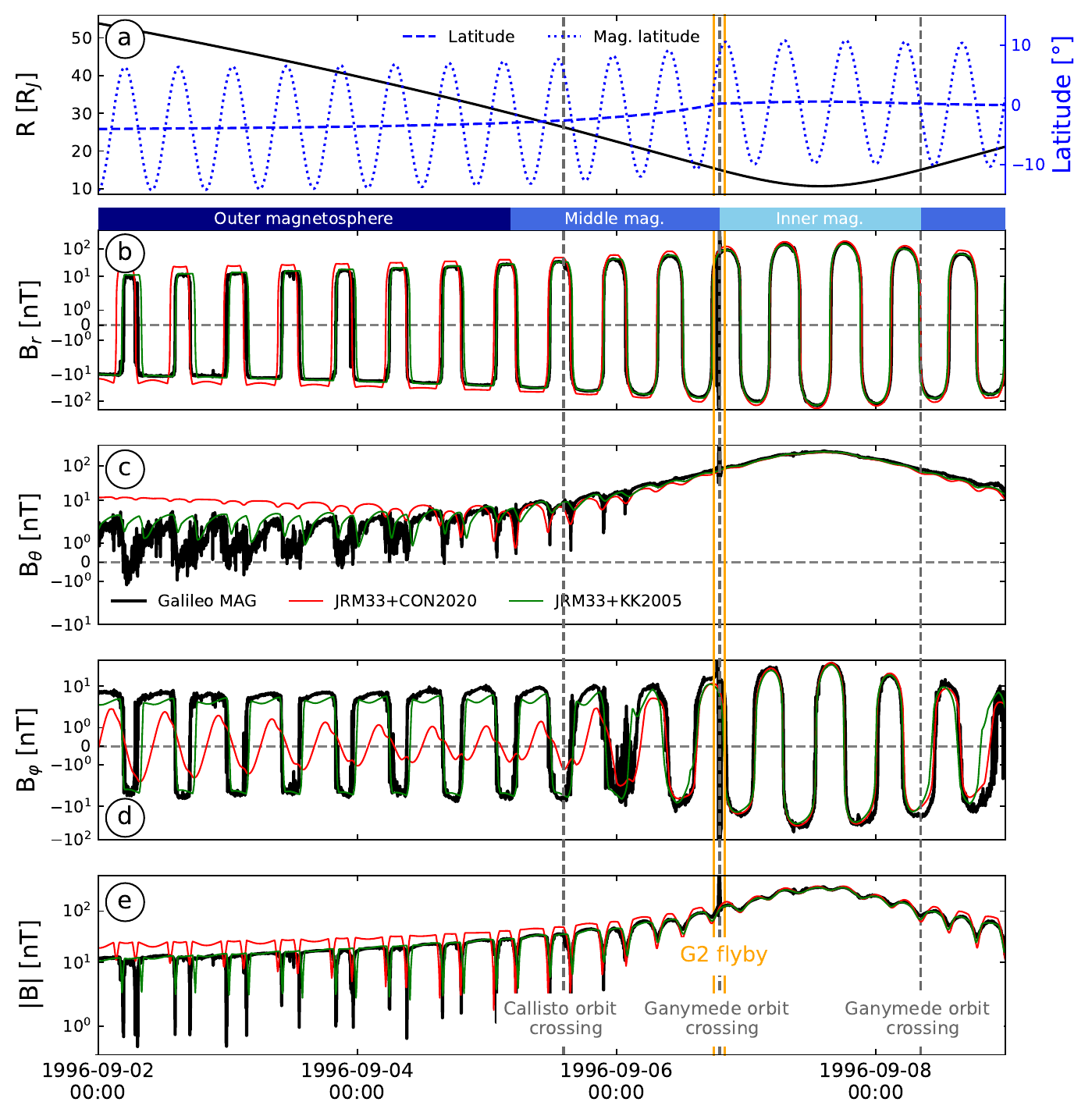}
    \caption{Comparisons of Galileo MAG data measured in the current sheet between 1996-09-02 and 1996-09-09 (Orbit G2) with the magnetic field components calculated by the JRM33+CON2020 (red curves) and JRM33+KK2005 (green curves) models. (a) Spacecraft radial distance (black line), jovigraphic latitude and magnetic latitude (blue dashed and dotted lines, respectively). Panels (b), (c), (d) and (e) show the different magnetic field components modeled and measured by the Galileo MAG instrument. Vertical dashed lines highlight the crossings of the Callisto and Ganymede orbits. The time interval corresponding to Galileo's G2 flyby of Ganymede is shown in orange.}
    \label{fig1}
\end{figure}

Figure \ref{fig1} shows a comparison between Galileo magnetic field measurements \cite{kivelson_1992} taken in the Jovian current sheet and the magnetic field components modeled by the JRM33+KK2005 and JRM33+CON2020 models. It should be noted that the time period chosen here includes a close flyby of Ganymede with a crossing of its own mini-magnetosphere which is not accounted for in the modeling results. The magnetic field components shown are calculated in the System-III Right-Handed coordinates system (S3RH), defined as the X-axis pointing in the Jupiter-Earth direction on January 1$^{\mathrm{st}}$, 1965 at 00:00 UT, the Z-axis as the planet's spin axis, and the Y-axis completing the trihedron \cite{dessler_1983}. This	system	rotates	with the planet's spin period of 9.92492 hours. We find that in the outer magnetosphere (R $>$ 30 $\textit{R}_{\mathrm{J}}$), the inversion of the radial and azimuthal magnetic field direction are accurately reproduced by both models (Figures \ref{fig1}b and \ref{fig1}d). The amplitude of the radial magnetic field is well estimated by both models (Figure \ref{fig1}b)  but the models show significant differences for the B$_{\theta}$ and B$_{\phi}$ components (Figures \ref{fig1}c and \ref{fig1}d). Indeed, in the outer region of the magnetosphere, the co-latitudinal component is overestimated by JRM33+CON2020 by a factor of 5 whereas the azimuthal one is underestimated by a factor of 3, while the data-model difference with JRM33+KK2005 is lower than 30$\%$.  As a direct consequence, the total magnetic field strength is better fitted with JRM33+KK2005 (Figure \ref{fig1}e). In the middle magnetosphere (15 $\textit{R}_{\mathrm{J}}$  $<$ R $<$ 30 $\textit{R}_{\mathrm{J}}$), the discrepancies between the two models and the observations reduces with decreasing radial distance, until in the inner magnetosphere (R $<$ 15 $\textit{R}_{\mathrm{J}}$), where both models are in good agreement with the Galileo data for all magnetic field components. We note that the similarity between the KK2005 and CAN81 models in the inner regions of the magnetosphere has already been identified in radiation belt studies \cite{sicard_2004}.

\begin{figure}[h!]
    \centering
    \includegraphics[width=0.85\textwidth]{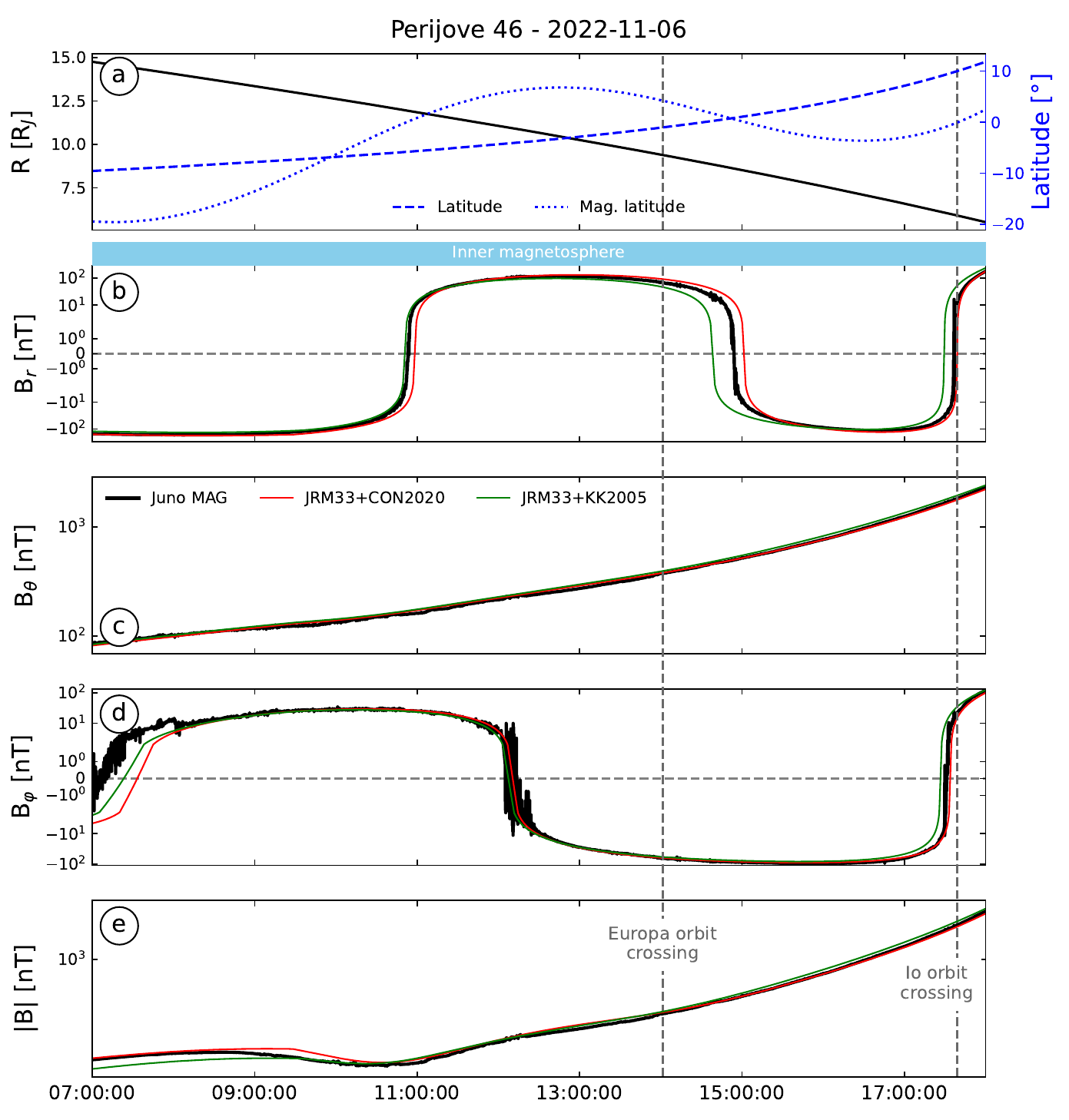}
    \caption{Comparison of Juno MAG data measured during a current sheet crossing and the model estimates. The panels displays the same physical parameters as those in Figure 1. Vertical dashed lines highlight the crossing of the orbit of Europa and Io.}
    \label{fig2}
\end{figure}

\clearpage

The orbit of Juno and its magnetic field instrumentation \cite{connerney_2017} enable to contrast data and models at higher latitude than Galileo and in the innermost part of the magnetosphere. Figure 2 presents Juno magnetic field measurements acquired during perijove 46 (perijove is used to count orbit number), in sector of radial distance not explored in Figure \ref{fig1}, i.e., R $<$ 12 $\textit{R}_{\mathrm{J}}$. At this time, a few hours before arriving above Jupiter’s North pole, Juno lies within the Jovian current sheet at R $\approx$ 8 $\textit{R}_{\mathrm{J}}$, i.e., between the orbits of Io (5.9 $\textit{R}_{\mathrm{J}}$) and Europa (9.4 $\textit{R}_{\mathrm{J}}$). Figure \ref{fig2} shows that, in this region, the magnetic field amplitudes are well reproduced by both models. However, we note that there is a few minutes offset between the radial magnetic field inversion measurements observed by Juno and the JRM33+KK2005 estimates while the change of sign of the azimuthal magnetic field component is well reproduced. The respective inversions are predicted with an accuracy of a few seconds using the JRM33+CON2020 model.      

\section{Comparison of the two current sheet models against UV observations of the Io, Europa, and Ganymede footprints}

The models can also be used to compute the magnetic field vector all along the flux tube that connects a moon to the Jovian atmosphere. In this article, magnetic field lines are computed using a constant step size of 1/350 $\textit{R}_{\mathrm{J}}$, which is approximately 200 km.  We note that using an adaptive step size given by ds = 10$^{-2} \times$ (log$_{10}$B)$^{-1}$, with B the magnetic field strength at the point of calculation, or a smaller constant step size leads to marginal differences of the order of 0.1$^{\circ}$ in footprint position estimates, but significantly increases calculation time.

We now focus on the inner magnetospheric region where Io, Europa, and Ganymede are connected to the atmosphere of Jupiter and compare the predictions of both models with observations of the moon auroral footprint positions. By doing so, we compare the validity of the models not only near the equator as done on Figure \ref{fig1} and \ref{fig2} with Galileo and Juno data, but also at mid and high latitudes all along the moon flux tubes. We first use both models to estimate Juno's high-latitude crossings of moons' flux tubes. We then compare the observed positions of the auroral spots linked to the moons with our predictions, which corresponds to a check of the models' accuracy at even higher latitudes, i.e., above the poles of Jupiter. We note that the CON2020 has already been validated against the timing of Europa and Ganymede flux tube crossings by Juno \cite<see Figures 7 and 8 of>{connerney_2020}. We also point out that our study is the first to  compare the position of the moons’ auroral UV footprints against the predictions of the KK2005 current sheet model.     
Using both current sheet models coupled to the JRM33 internal magnetic field model, we apply a field-line tracing method in order to map magnetically the position of the moon auroral footprints onto Jupiter’s atmosphere. The moons’ footprints are computed for an altitude of 900 km above Jupiter’s surface, which corresponds to the altitude of the moon-induced UV aurora \cite{bonfond_2009}. Jupiter’s surface is defined as an ellipsoid whose polar and equatorial axis are 66,854 km and 71,492 km, respectively. 

\begin{figure}[h!]
    \centering
    \includegraphics[width=\textwidth]{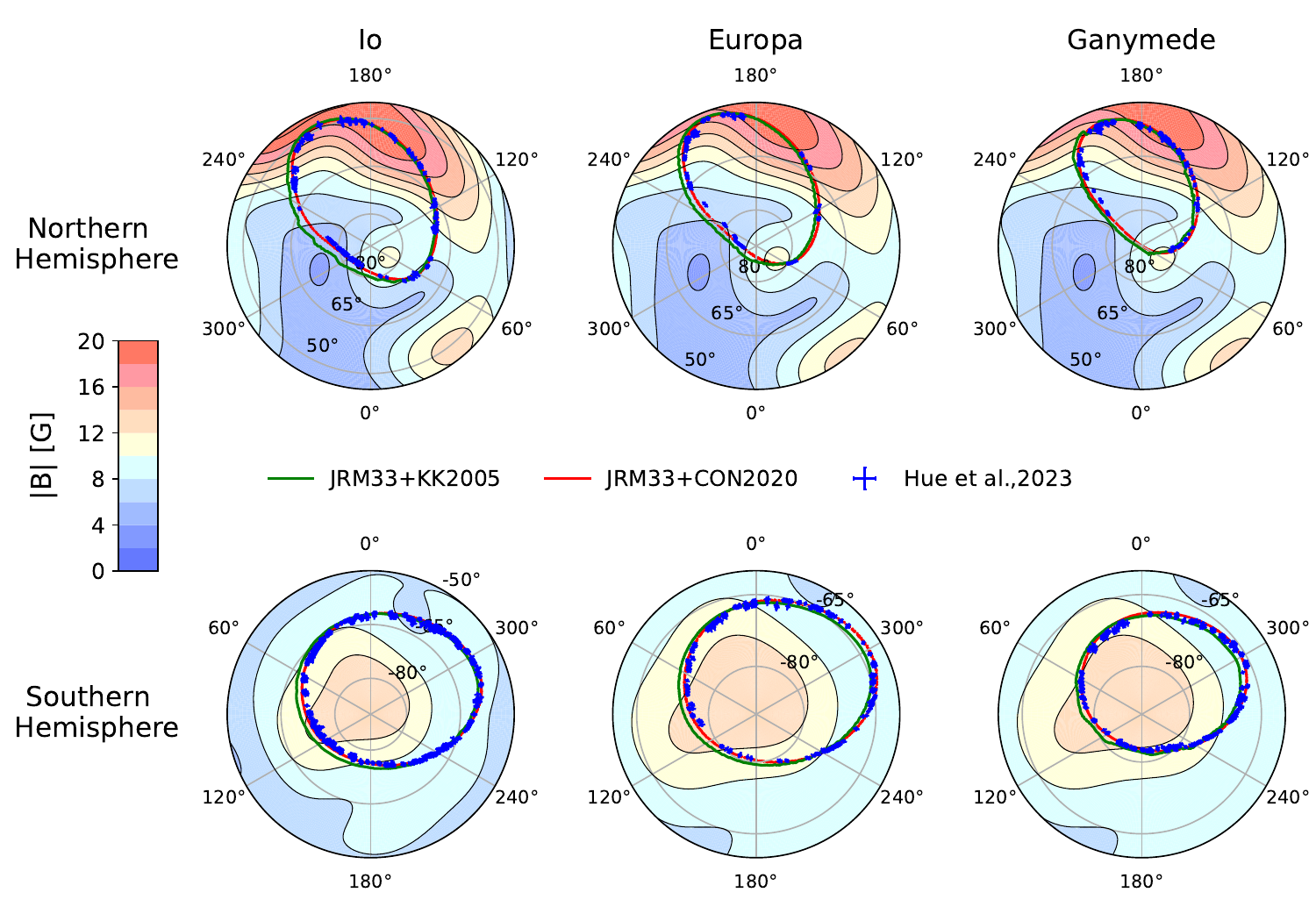}
    \caption{Positions of the footprints of Io (left), Europa (center) and Ganymede (right) in the northern (top) and southern (bottom) hemispheres. The footprints positions are derived using the two current sheet models coupled with JRM33 and compared to the true positions observed and documented in \citeA{hue_2023}. The magnetic field intensity at the 1-bar level, calculated with JRM33, is displayed in the background.}
    \label{fig3}
\end{figure}

We compare our calculations with the positions of UV spots of the moon as reported in \citeA{hue_2023}. This last study, which includes a large number of observations of Io, Europa, and Ganymede footprints made by the UVS instrument \cite{gladstone_2017} onboard the Juno spacecraft \cite{bolton_2017} during its 43 perijoves, provides an unprecedented capability to assess the accuracy of magnetic field models for the magnetic mapping of moons’ UV spots. We did not include in our study the only probable candidate for the Callisto UV footprint reported by \citeA{bhattacharyya_2018}. 

We calculate the positions of the footprints of Io, Europa, and Ganymede in the northern and southern hemispheres considering 20 of their respective orbital periods. This corresponds to 36 days for Io, 71 days for Europa and 143 days for Ganymede. The starting date used to calculate positions and times has been arbitrarily set at 2018-01-01. These time periods provide sufficient spatial sampling to be representative of the variations induced by the local time. Finally, we average the data using a sliding window over a 20-degree longitude interval in order to remove the smallest variations. The results of our calculations are displayed in Figure 3. The System-III Right Handed coordinates are used for representation. We note that only the KK2005 model can include the variability of the Jovian magnetosphere with local time, and that in our case the variability induced by the local time appears as a wobbling of the moon auroral footpaths. This effect is more prominent in the northern than in the southern hemisphere of Jupiter.  

On the North pole, in the 60$^{\circ}$ - 220$^{\circ}$ longitude sector, the two model estimates overlap and match observations of the moons’ footprints. The major difference between the moon auroral footpaths computed with the two current sheet models and the observations is located in the 220$^{\circ}$ - 60$^{\circ}$ area. Indeed, in this sector, the moon auroral footprints derived with JRM33 + KK2005 are more extended southward than the other ones. We note that the currently published Juno UVS observations do not provide a complete longitudinal coverage of the moon footpaths, so more UV observations in the future would help assess the accuracy of the model. 

On the South pole, the footpaths derived using the two models have almost exactly the same shape. However, that of JRM33+KK2005 is shifted in the northward direction towards longitude $\lambda_{\mathrm{S3RH}}$ = 110$^{\circ}$. As a result, the footpaths of the two models do not overlap in almost all longitude sectors. We find that the Juno UVS observations are closer to the JRM33+CON2020 estimates than to the JRM33+KK2005 estimates.

\begin{figure}[h!]
    \centering
    \includegraphics[width=\textwidth]{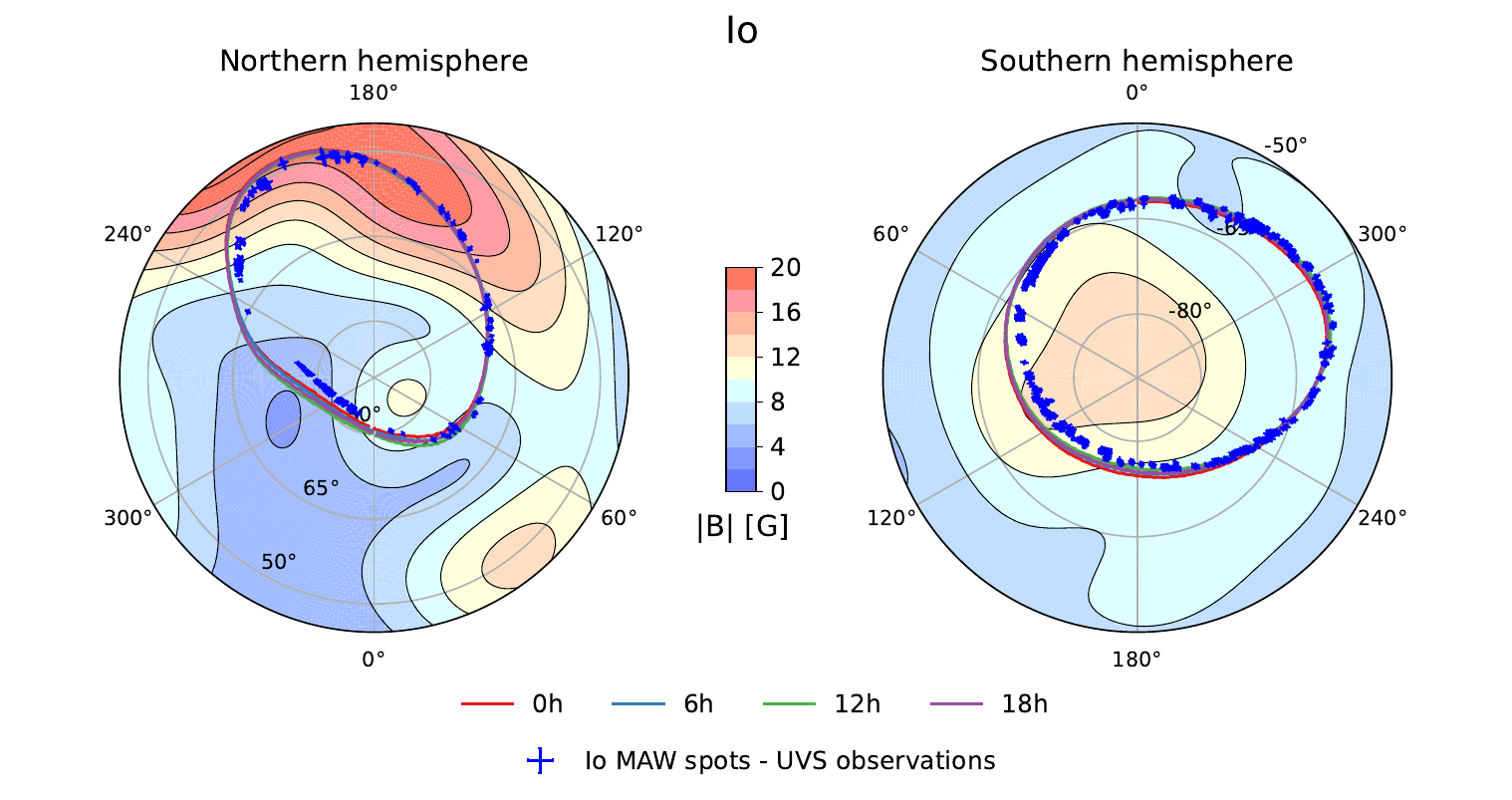}
    \caption{Positions of the Io footprints as a function of the moon local time derived with JRM33+KK2005. Calculations are made by setting Io at four constant local time along its orbit: 0h (Midnight ; Red curve), 6h (Dawn ; Blue curve), 12h (Noon ; Green curve), 18h (Dusk ; Purple curve). UVS observations of the Io MAW spots are shown for comparison with the actual positions.}
    \label{fig4}
\end{figure}

In order to illustrate the effect of local time on the footprint positions calculated with KK2005, we derive Io’s footprints by setting the moon at a constant local time. Calculations are done for local times from midnight (LT = 0h) to dusk (LT = 18h), and over a time interval of 20 orbital periods of the moon, as used in the previous calculation. The results, displayed in Figure \ref{fig4}, show that for a constant local time, the footpath is continuous and exhibits no small-scale oscillations. However, the four footpaths corresponding to four different local times diverge mainly in the 240$^{\circ}$ - 0$^{\circ}$ sector of the northern hemisphere. This is the reason why the footpaths displayed in Figure \ref{fig3}, derived from the actual position of the moons along their orbits and therefore from a succession of different local times, exhibit an oscillation in this region. The local time effects on the footprints position predicted here with the KK2005 model have not been observed in the Juno UVS data, or have not been shown with the limited UVS data currently available. However, modulation of the UV power emitted by the Io footprint as a function of local time has been previously reported \cite{hue_2019}.  

\begin{figure}[h!]
    \centering
    \includegraphics[width=\textwidth]{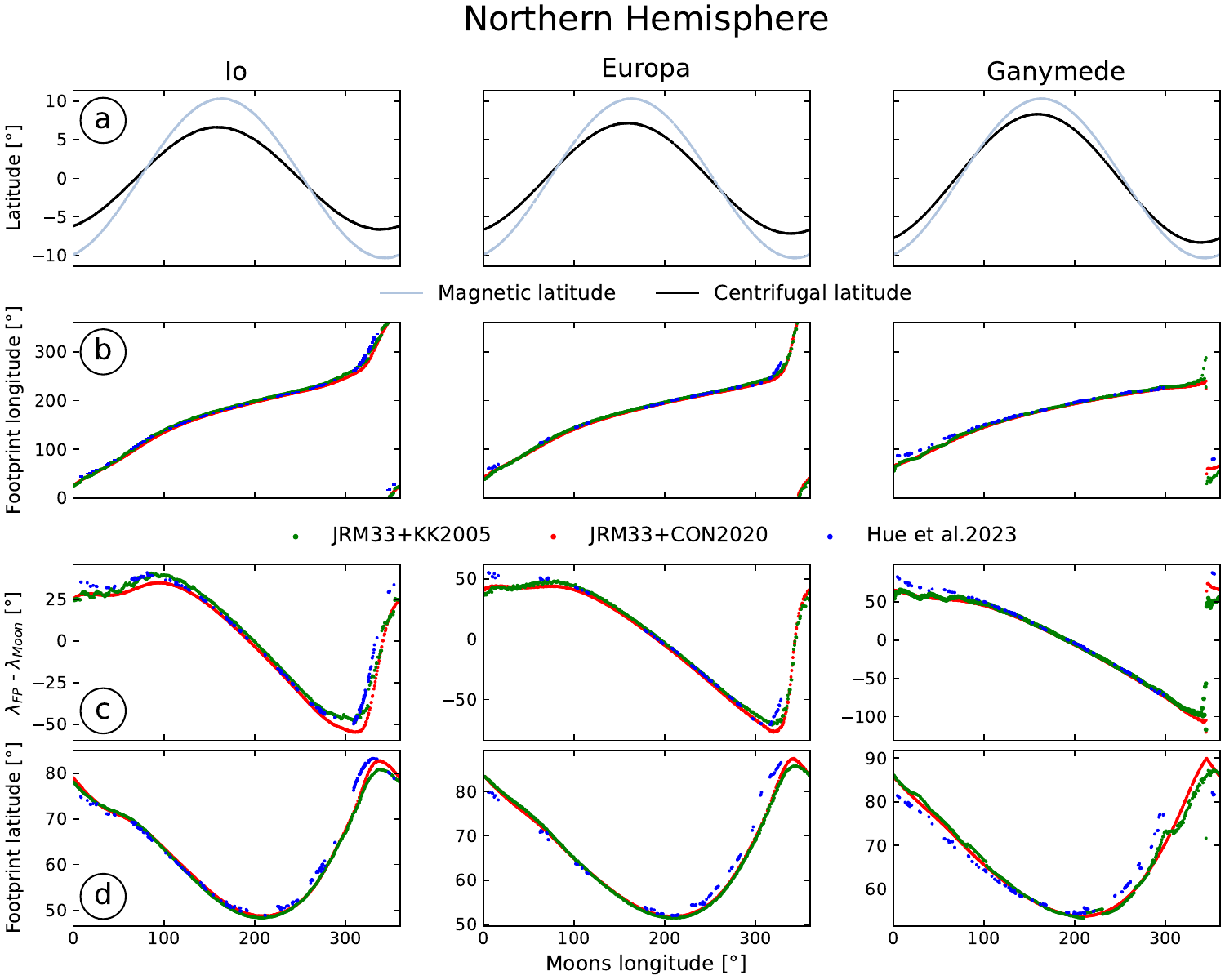}
    \caption{(a) Magnetic and centrifugal latitudes of Io, Europa and Ganymede as a function of their longitudes. (b) Longitudes of the magnetic footprints in the northern hemisphere calculated with JRM33+CON2020 (red), JRM33+KK2005 (green) and reported by Hue et al., (2023) (blue). Color code is the same for panels (b), (c) and (d). (c) Longitudinal separation between the magnetic footprints and the moons in the northern hemisphere. (d) Footprints’ latitudes. }
    \label{fig5}
\end{figure}

\begin{figure}[h!]
    \centering
    \includegraphics[width=\textwidth]{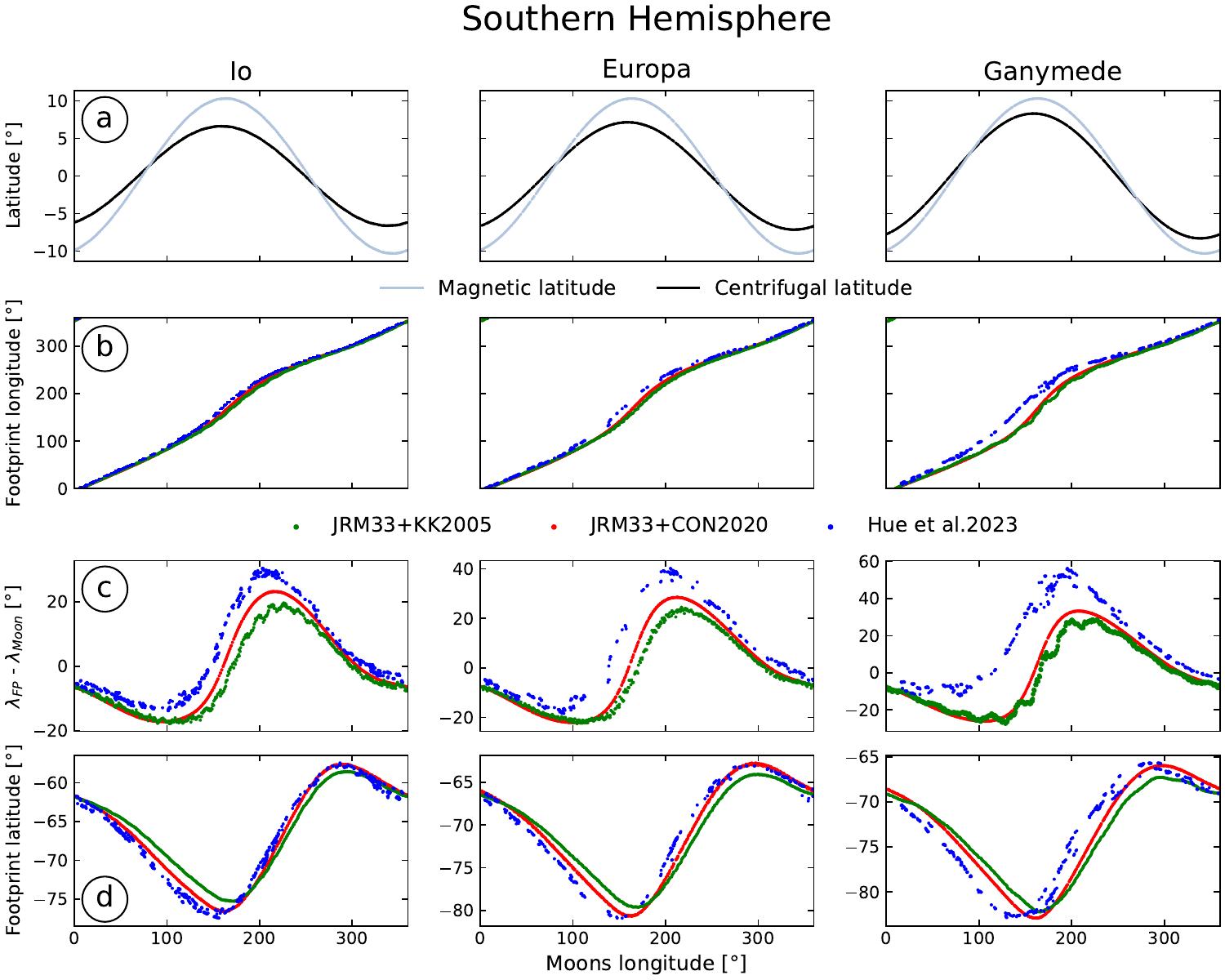}
    \caption{Same as Figure 5, for the southern hemisphere.}
    \label{fig6}
\end{figure}

Figures \ref{fig5} and \ref{fig6} show the results of the calculation displayed in Figure \ref{fig3} in rectangular graphs, function of moons’ System-III Right-Handed longitudes, for both the northern and southern hemispheres, respectively. We note that the two models reproduce the global trend of UV auroral observations. In the northern hemisphere, we observe that the footprint positions are accurately reproduced by both models (Figure \ref{fig5}b and \ref{fig5}d). However, the discrepancies between models and observations increase as we move from Io to Europa to Ganymede. In the southern hemisphere, this difference is much greater. The footprints positions are better estimated with JRM33+CON2020 although we also note increasing errors from Io to Ganymede (Figure \ref{fig6}b and \ref{fig6}d). The footpaths derived using JRM33+KK2005 have a shape and amplitude close to what has been observed but are shifted by a few tens of latitudinal degrees towards increasing longitudes (Figure \ref{fig6}d). This offset already exists in the footpaths displayed in Figure \ref{fig3}. In both hemispheres, an oscillation of the JRM33+KK2005 estimates due the consideration of the local time explored by the moons are observed, as shown in Figure \ref{fig3}. 

Figure \ref{fig5}c and \ref{fig6}c contrast models and UV data against longitudinal separations between the moon and its footprint as a function of moons’ longitude, thus indicating the longitudinal bending of the magnetic field lines between the orbital planes of the moons and Jupiter’s poles. These differences are linked to the physical properties of the charged particles that generate the footprints. The MAW spots are created by electrons traveling at finite speeds along the magnetic field lines from the vicinity of the moons to Jupiter’s ionosphere. In contrast, the positions derived using the models are instantaneous footprints that assume infinite propagation speed and no local deformations of the magnetic field lines. Thus, the longitudinal separations between the MAW spots and the footprints calculated with the models are due to the local deformation of the magnetic field lines close to the moons and to the finite propagation speed of the particle.

The lead angles are greater when the particles have to pass through the current sheet \cite{hess_2010,louis_2019,hue_2023}. For example, in Figure \ref{fig5} for the northern hemisphere, the largest differences between simulations and observations are located in the 254$^{\circ}$ - 71$^{\circ}$ longitude sector, when the moons are located below the magnetic equator. Particles precipitating towards the north pole therefore have a much longer travel time, and thus a greater lead angle due to the travel time, causing a discrepancy between modeling and observations. The same phenomenon is observed in Figure \ref{fig6} for the southern hemisphere, in the 71$^{\circ}$ - 254$^{\circ}$ longitude sector when the moons are above the current sheet. 

\section{Comparison of the two current sheet models against Juno in situ observations of the Io, Europa, and Ganymede flux tube crossings} 

Figures \ref{fig3}, \ref{fig4}, \ref{fig5} and \ref{fig6} compare the current sheet models with the Juno-UVS observations and their ability to map the magnetic footprints of Io, Europa and Ganymede on Jupiter’s poles. We now track the magnetic field lines to the equatorial part of the magnetosphere. Using the two models, we derive the corresponding M-Shell parameter by following the magnetic field line down to the magnetic equator, i.e., the position where the magnetic field strength is the weakest along the same magnetic field line. The M-Shell parameter is the distance from the center of Jupiter of the point where the magnetic field line crosses the magnetic equator, normalized to the Jovian radius. The wording "equatorial projection" is used in this article to refer to the radial distance where a magnetic field line crosses the orbital planes of the moons, i.e., when the planetographic latitude is zero.

\begin{figure}[h!]
    \centering
    \includegraphics[width=\textwidth]{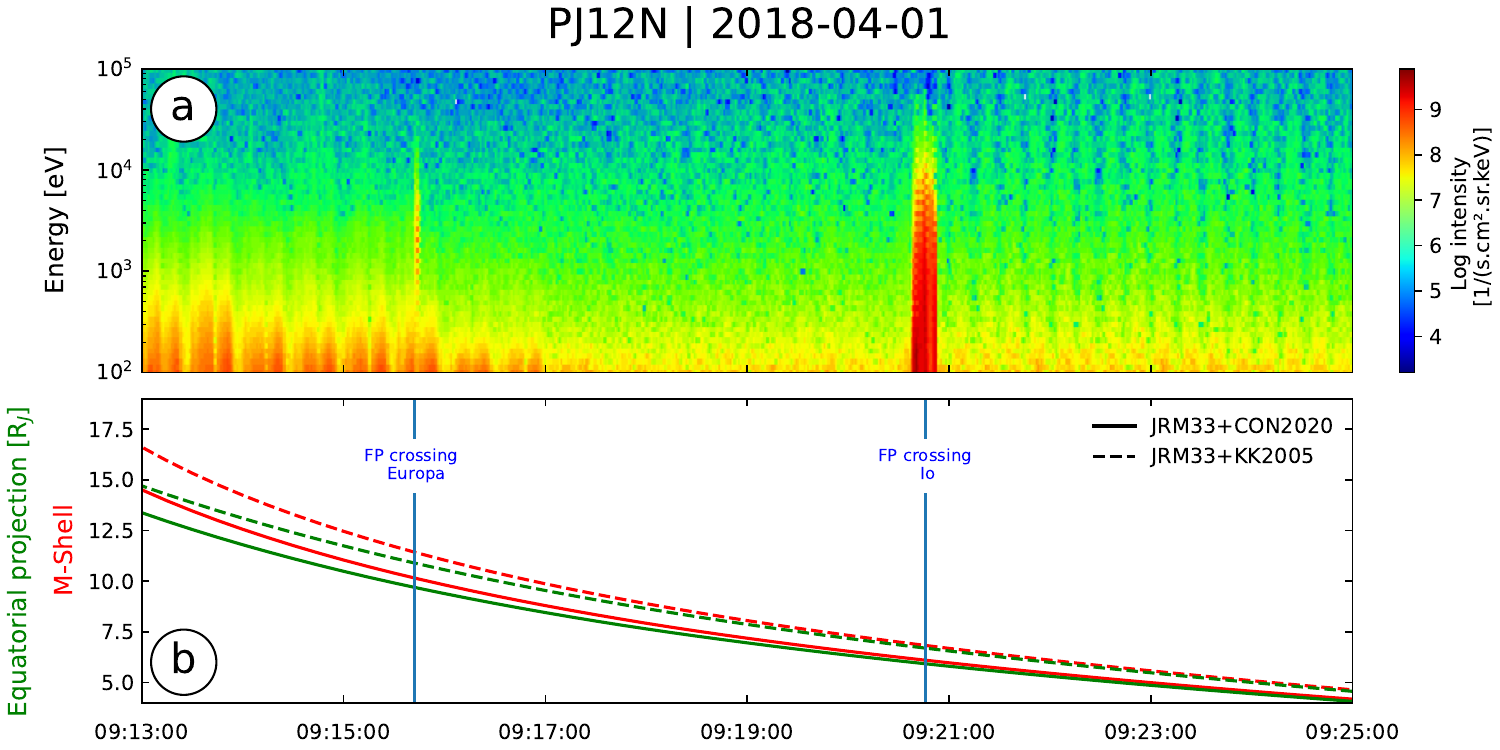}
    \caption{(a) Time-Energy-Differential Number Flux (DNF) spectrogram of the JADE-E data recorded during Juno’s 12th perijove. (b) Magnetic projection of Juno’s position onto the orbital plane of the Galilean moons (green curves) and Juno’s M-Shell (red curves) calculated with JRM33+KK2005 (dashed lines) and JRM33+CON2020 (plain lines). Vertical blue lines at 09:15:42 and 09:20:46 highlight the magnetic flux tubes crossings of Europa and Io, respectively. Juno's position and velocity with respect to Jupiter are indicated below the sub-panel (b).}
    \label{fig7}
\end{figure}

Figure \ref{fig7} shows Juno’s M-Shell and equatorial projection during PJ12 and compares them to the JADE-E data. The perijove 12 north was chosen to compute the two parameters because the JADE-E instruments \cite{mccomas_2017} consecutively measured two enhancements in the electron differential number flux associated with Europa’s and Io’s flux tube crossings \cite{allegrini_2020,szalay_2020a}. We find that the M-Shell and the equatorial projections derived with the JRM33+KK2005 are always higher than those calculated with JRM33+CON2020 on this time range.  The Juno’s equatorial projections made with JRM33+KK2005 during the crossings of Europa’s (09:15:42) and Io’s flux tubes (09:20:46) give 10.90 $\textit{R}_{\mathrm{J}}$ and 6.69 $\textit{R}_{\mathrm{J}}$, respectively. These results are slightly higher than expected given that r$_{\mathrm{Eur}}$ and r$_{\mathrm{Io}}$, the radial distances covered by Europa and Io, are 9.38 $\pm$ 0.08 $\textit{R}_{\mathrm{J}}$ and 5.90 $\pm$ 0.03 $\textit{R}_{\mathrm{J}}$, respectively. The results obtained with JRM33+CON2020 of 9.72 $\textit{R}_{\mathrm{J}}$ and 5.93 $\textit{R}_{\mathrm{J}}$, respectively, are closer to the actual orbital distances of the moons. 

JRM33+KK2005 estimates of Juno’s M-Shell during the moon flux tube crossings are in good agreement for the Io flux tube crossing (M = 6.84) but also slightly higher than expected for the Europa flux tube crossing (M = 11.45). With JRM33+CON2020, we found Juno M-Shell values of 10.17 and 6.1 for the Europa and Io flux tube crossings, respectively, which are within the range of values covered by each moon \cite{paranicas_2018}. 

Finally, we examine the accuracy of the two current sheet models to predict the crossings of moons’ flux tubes. To do so, we compute the magnetic footpath of both Juno and the moon onto the Jovian atmosphere, and then we identify the timing of the intersection between the two projected trajectories. We compare our estimates with Juno crossings of moons’ flux tubes reported in various studies. Statistical studies using in situ measurements within the Io flux tubes have been performed by \citeA{szalay_2018}, \citeA{szalay_2020a},  \citeA{clark_2023}, and \citeA{sulaiman_2023}. Tables detailing the identified timings of their observations  are provided in the Supplementary Information files of these articles. We note that a comparison of the JRM33+CON2020 estimates with Io flux tubes crossings observations is already included in the supplementary material of \citeA{clark_2023}. We therefore use the estimates provided in this study.  Crossings of Europa flux tubes have been reported in \citeA{allegrini_2020} and \citeA{rabia_2023}. Finally, for Ganymede, two events have been documented. The first one, a crossing of a flux tube connected to the auroral tail, was studied by \citeA{szalay_2020b} and \citeA{louis_2020}. A TEB crossing is analyzed in \citeA{hue_2022}. As we calculate the beginning of the flux tube crossings with the two models, we compare our estimates with the start times of the observations provided in the previously cited articles. The results of the calculation are provided in Table 1.

\begin{table}[h!]
\scriptsize
\makebox[\linewidth]{
\hspace{-1cm}
\begin{tabular}{c|c|c|c|c|c|c}
\hline
 Moon &
  PJ &
  Date &
  Moon LT [h] &
  \begin{tabular}[c]{@{}c@{}}Observed time in   \\ JADE/JEDI data\end{tabular} &
  \begin{tabular}[c]{@{}c@{}}Predicted time   \\ JRM33+CON20\end{tabular} &
  \begin{tabular}[c]{@{}c@{}}Predicted time  \\ JRM33+KK2005\end{tabular} \\
\hline
\multirow{43}{*}{Io}      & 1S$^{1}$      & 2016-08-27 & 23.4 & 13:20:35            & 13:20:45  \textcolor{green}{(+10)}       & 13:21:14  \textcolor{red}{(+39)} \\
                          & 3N$^{1}$      & 2016-12-11 & 22.7 & 16:32:35            & 16:32:41  \textcolor{green}{(+06)}       & 16:32:22   \textcolor{red}{(-13)}  \\
                          & 3S$^{1}$      & 2016-12-11 & 23.3 & 17:30:43            & 17:31:17  \textcolor{green}{(+34)}       & 17:31:28  \textcolor{red}{(+45)}  \\
                          & 4S$^{1}$      & 2017-02-02 & 19.7 & 13:29:17            & 13:29:51  \textcolor{red}{(+34)}       & 13:29:00  \textcolor{green}{(-17)}   \\
                          & 5N$^{1,2,3,4}$  & 2017-03-27 & 15.7 & 08:34:37 / 08:34:36 & 08:34:43  \textcolor{red}{(+6)/(+5)}   & 08:34:40  \textcolor{green}{(+3)/(+4)} \\
                          & 5S$^{2,3,4}$    & 2017-03-27 & 16.2 & 09:30:51            & 09:31:00 \textcolor{green}{(+9)}         & 09:29:56  \textcolor{red}{(-55)}  \\
                          & 6N$^{1,2,4}$   & 2017-05-19 & 12.1 & 05:39:35 / 05:39:28 & 05:39:57  \textcolor{green}{(+22)/(+29)} & 05:40:10  \textcolor{red}{(+35)/(+42)} \\
                          & 6S$^{1,2,3,4}$  & 2017-05-19 & 13.3 & 06:40:04 / 06:39:53 & 06:39:50  \textcolor{green}{(-14)/(-3)}  & 06:39:24  \textcolor{red}{(-40)/(-29)}  \\
                          & 7N$^{1,2,4}$    & 2017-07-11 & 9.1  & 01:24:52 / 01:24:40 & 01:24:58  \textcolor{green}{(+6)/(+18)}  & 01:25:32  \textcolor{red}{(+40)/(+52)} \\
                          & 7S$^{1,2,3,4}$  & 2017-07-11 & 9.6  & 02:22:33 / 02:22:27 & 02:22:27  \textcolor{green}{(-6)/(=)}    & 02:23:30  \textcolor{red}{(+57)/(+63)} \\
                          & 8N$^{1}$      & 2017-09-01 & 5.5  & 21:23:09            & 21:23:55  \textcolor{red}{(+46)}       & 21:22:58  \textcolor{green}{(-11)} \\
                          & 9N$^{1}$      & 2017-10-24 & 2.0  & 17:22:55            & 17:22:57  \textcolor{green}{(+02)}       & 17:22:44  \textcolor{red}{(-11)} \\
                          & 10S$^{1}$     & 2017-12-16 & 1.3  & 18:30:50            & 18:30:48  \textcolor{green}{(-02)}       & 18:30:46  \textcolor{red}{(-04)}   \\
                          & 12N$^{1,2,3}$ & 2018-04-01 & 17.6 & 09:20:42 / 09:20:37 & 09:20:56  \textcolor{green}{(+14)/(+19)} & 09:22:10  \textcolor{red}{(+88)/(+93)} \\
                          & 12S$^{1}$     & 2018-04-01 & 18.2 & 10:28:04            & 10:28:26  \textcolor{red}{(+22)}       & 10:28:24  \textcolor{green}{(+20)} \\
                          & 13N$^{1,2,3}$ & 2018-05-24 & 14   & 05:13:33 / 05:13:11 & 05:13:39  \textcolor{green}{(+06)}/\textcolor{red}{(+28)} & 05:13:18  \textcolor{red}{(-15)}/\textcolor{green}{(+7)}  \\
                          & 13S$^{1}$     & 2018-05-24 & 14.5 & 06:10:59            & 06:11:25  \textcolor{green}{(+26)}       & 06:12:04  \textcolor{red}{(+65)} \\
                          & 14N$^{1,2,3}$ & 2018-07-16 & 12.5 & 04:50:01 / 04:49:50 & 04:50:19  \textcolor{green}{(+18)/(+29)} & 04:50:56  \textcolor{red}{(+55)/(+66)}  \\
                          & 14S$^{1}$     & 2018-07-16 & 13.1 & 05:53:41            & 05:53:27  \textcolor{green}{(-14)}       & 05:54:44  \textcolor{red}{(+63)}   \\
                          & 15N$^{1,2}$   & 2018-09-07 & 8.9  & 00:48:39 / 00:48:14 & 00:49:01  \textcolor{red}{(+22)/(+47)} & 00:48:18  \textcolor{green}{(-21)/(+4)} \\
                          & 15S$^{1,2,3}$ & 2018-09-07 & 9.5  & 01:45:55 / 01:46:00 & 01:46:21  \textcolor{green}{(+26)/(+21)} & 01:46:26  \textcolor{red}{(+31)/(+26)}  \\
                          & 16N$^{1,2}$   & 2019-10-29 & 11.0 & 20:47:48 / 20:47:38 & 20:47:34  \textcolor{green}{(-14)/(-4)}  & 20:47:30  \textcolor{red}{(-18)/(-8)}  \\
                          & 16S$^{2,3}$   & 2019-10-29 & 11.6 & 21:47:20            & 21:46:56  \textcolor{green}{(+24)}       & 21:45:54 \textcolor{red}{(-86)} \\
                          & 18N$^{1}$     & 2019-02-12 & 0.9  & 17:18:09            & 17:17:55  \textcolor{green}{(-14)}       & 17:17:48  \textcolor{red}{(-21)}  \\
                          & 20S$^{1}$     & 2019-05-29 & 17.5 & 08:46:27            & 08:46:05  \textcolor{green}{(-22)}       & 08:47:04  \textcolor{red}{(+37)} \\
                          & 21S$^{1}$     & 2019-07-21 & 13.9 & 04:45:53            & 04:45:43  \textcolor{green}{(-10)}       & 04:45:34  \textcolor{red}{(-19)} \\
                          & 22N$^{1,2,3}$ & 2019-09-12 & 11.8 & 03:18:55 / 03:18:52 & 03:19:29  \textcolor{red}{(+34)/(+37)} & 03:18:42  \textcolor{green}{(-13)/(-10)}  \\
                          & 23N$^{2}$     & 2019-11-03 & 7.6  & 22:07:02            & 22:07:08 \textcolor{green}{(+6)}         & 22:07:14 \textcolor{red}{(+12)}  \\
                          & 23S$^{1,2,3}$ & 2019-11-03 & 8.2  & 23:10:48 / 23:10:38 & 23:10:42  \textcolor{green}{(-6)/(+4)}   & 23:09:28  \textcolor{red}{(-80)/(-70)}  \\
                          & 24N$^{1}$     & 2019-12-26 & 3.5  & 17:09:33            & 17:09:19  \textcolor{red}{(-14)}       & 17:09:44  \textcolor{green}{(+11)} \\
                          & 24S$^{2}$     & 2019-12-26 & 4.2  & 18:30:16            & 18:30:32  \textcolor{red}{(+16)}       & 18:30:20  \textcolor{green}{(+4)} \\
                          & 26N$^{1}$     & 2020-04-10 & 22.7 & 13:23:12            & 13:23:18  \textcolor{red}{(+06)}       & 13:23:08  \textcolor{green}{(-04)}  \\
                          & 26S$^{2}$     & 2020-04-10 & 23.4 & 14:34:57            & 14:33:48  \textcolor{red}{(-69)}        & 14:34:32 \textcolor{green}{(-25)} \\
                          & 27N$^{1}$     & 2020-06-02 & 19.5 & 09:59:35            & 10:00:13  \textcolor{red}{(+38)}        & 09:59:16  \textcolor{green}{(-19)}  \\
                          & 28N$^{1}$     & 2020-07-25 & 16.0 & 06:00:09            & 05:59:51 \textcolor{green}{(-18)}        & 05:59:46  \textcolor{red}{(-23)}   \\
                          & 28S$^{1}$     & 2020-07-25 & 16.5 & 07:02:13            & 07:01:59  \textcolor{green}{(-14)}       & 07:01:30  \textcolor{red}{(-43)}  \\
                          & 29S$^{1,3}$   & 2020-09-16 & 12.4 & 02:00:35 / 02:00:30 & 02:00:29  \textcolor{green}{(-6)/(-1)}   & 02:01:00  \textcolor{red}{(+25)/(+30)}  \\
                          & 30S$^{1}$     & 2020-11-08 & 11.5 & 02:41:56            & 02:42:18  \textcolor{green}{(+22)}       & 02:40:58  \textcolor{red}{(-58)}  \\
                          & 31S$^{1}$     & 2020-12-30 & 7.9  & 22:45:07            & 22:45:25  \textcolor{green}{(+18)}       & 22:44:40  \textcolor{red}{(-27)} \\
                          & 32N$^{1}$     & 2021-02-21 & 3.6  & 17:19:51            & 17:19:53  \textcolor{green}{(+02)}       & 17:19:48  \textcolor{red}{(-03)}  \\
                          & 32S$^{1}$     & 2021-02-21 & 4.2  & 18:23:24            & 18:22:26  \textcolor{red}{(-58)}        & 18:24:10  \textcolor{green}{(+46)} \\
                          & 33S$^{1}$     & 2021-04-16 & 6.2  & 00:18:00            & 00:17:58  \textcolor{green}{(-02)}       & 00:18:12  \textcolor{red}{(+12)}  \\
                          & 34N$^{1}$     & 2021-06-08 & 9.0  & 07:29:32            & 07:29:34 \textcolor{green}{(+02)}        & 07:29:06  \textcolor{red}{(-26)}  \\
\hline
\multirow{10}{*}{Europa}  & 5N$^{6}$      & 2017-06-08 & 19.7 & 08:32:36            & 08:32:42   \textcolor{red}{(+06)}      & 08:32:40  \textcolor{green}{(+04)}   \\
                          & 12N$^{5,6}$   & 2018-04-01 & 18.6 & 09:15:42            & 09:15:58  \textcolor{green}{(+16)}       & 09:16:42  \textcolor{red}{(+60)}  \\
                          & 13N$^{6}$     & 2018-05-24 & 15.3 & 05:09:38            & 05:09:34  \textcolor{green}{(-04)}       & 05:09:28  \textcolor{red}{(-10)} \\
                          & 15S$^{6}$     & 2018-09-07 & 10.2 & 01:50:10            & 01:50:06  \textcolor{green}{(-04)}       & 01:50:38  \textcolor{red}{(+28)} \\
                          & 23S$^{6}$     & 2019-11-03 & 10.1 & 23:18:43            & 23:18:44  \textcolor{green}{(+01)}       & 23:17:18  \textcolor{red}{(-85)} \\
                          & 29N$^{6}$     & 2020-09-16 & 15.4 & 01:44:55            & 01:44:34  \textcolor{green}{(-21)}       & 01:42:22  \textcolor{red}{(-153)} \\
                          & 32S$^{6}$    & 2021-02-21 & 7.0  & 18:31:21            & 18:31:24  \textcolor{green}{(+03)}      & 18:33:20  \textcolor{red}{(+119)} \\
                          & 33S$^{6}$     & 2021-04-16 & 6.5  & 00:24:35            & 00:24:14  \textcolor{red}{(-21)}       & 00:24:24  \textcolor{green}{(-11)}  \\
                          & 34S$^{6}$     & 2021-06-08 & 6.7  & 08:37:21            & 08:37:10  \textcolor{green}{(-11)}       & 08:37:36  \textcolor{red}{(+15)} \\
                          & 41S$^{6}$     & 2022-04-09 & 4.6  & 16:51:50            & 16:51:32  \textcolor{green}{(-18)}       & 16:52:46  \textcolor{red}{(+56)} \\
\hline
\multirow{2}{*}{Ganymede} & 20N$^{7,8}$   & 2019-05-29 & 13.6 & 07:37:14            & 07:37:40  \textcolor{red}{(+26)}        & 07:37:18  \textcolor{green}{(+22)}  \\
                          & 30S$^{9}$     & 2020-11-08 & 8.5  & 02:55:00            & 02:24:58  \textcolor{green}{(-2)}        & 02:54:32  \textcolor{red}{(-28)} \\
\hline
\end{tabular}
}

\caption{Comparison of predicted and observed times for Juno's moons flux tubes crossings. For some crossings, two different starting times were reported, depending on the particles studied. Times differences, in second, between estimates and observations are given in brackets. Color code indicates the best (green) and the poorest (red) estimations. Index in the perijove (PJ) column indicate the article in which the crossing is studied : $^{1}$\citeA{clark_2023} - $^{2}$\citeA{szalay_2020a} - $^{3}$\citeA{sulaiman_2023} - $^{4}$\citeA{szalay_2018} - $^{5}$\citeA{allegrini_2020} - $^{6}$\citeA{rabia_2023} - $^{7}$\citeA{szalay_2020b} - $^{8}$\citeA{louis_2020}  - $^{9}$\citeA{hue_2022}}
\end{table}

\clearpage

We find that the JRM33+CON2020 estimates are closer to the observations than the JRM33+KK2005 ones in 72 $\%$ of cases. The median deviation between estimates and observations are 14 seconds and 26 seconds for JRM33+CON2020 and JRM33+KK2005, respectively. As the Juno spacecraft moves very rapidly above the auroral regions ($\sim$ 50 km.s$^{-1}$, Figure 7), it can sweep up a range of M-Shells larger than 10 in less than 30 seconds, depending on the orbit configuration. Thus, a significant time discrepancy between the predictions and the actual observations of the flux tube crossings could lead to a very different interpretation of the origins of the charged particles measured in situ in the magnetosphere.

\begin{figure}[h!]
    \centering
    \includegraphics[width=0.75\textwidth]{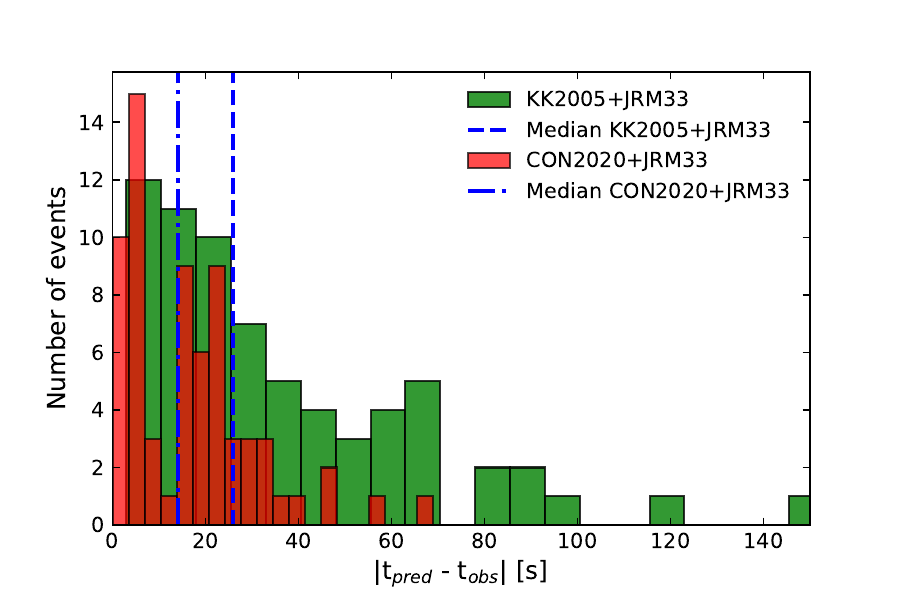}
    \caption{Distribution of the time difference between the flux tubes crossing observed by Juno  and the estimates using JRM33+CON2020 (red) and JRM33+KK2005 (green). Dashed  lines highlight the median deviations of each model  estimates.}
    \label{fig8}
\end{figure}

By taking into account only Io’s flux tube crossings, we find the same median deviations, as Table 1 mainly contains observations concerning Io. For Europa, median deviations are 8.5 seconds and 42 seconds for the JRM33+CON2020 and JRM33+KK2005 estimates, respectively. This result shows the accuracy of the estimates does not depend only on the radial distances. Considering the crossings located above the north and south poles separately, we found a discrepancy in the estimates in both hemispheres. Indeed, in the northern hemisphere, 65$\%$ of the closest estimates are made using JRM33+CON2020 while in the southern hemisphere, this model provided 80$\%$ of the best estimates. The median deviation of JRM33+CON2020 estimates is 15 and 14 seconds for the crossings in the northern and southern hemispheres, respectively. However, we point out that JRM33+KK2005 estimates show a significant north/south discrepancy. Indeed, in the southern hemisphere, the median deviation between the time of the crossings and the estimates falls to 37 seconds while it is 16.5 seconds in the northern hemisphere. These values are consistent with the shift between the JRM33+KK2005 footpaths in the southern hemisphere and the UV observations, as shown in Figure \ref{fig3}. Indeed, the footpaths do not fit the actual positions of the MAW spots over a wide range of longitude while in the northern hemisphere, the gap between the Juno UVS data and KK2005 estimates is mainly observed in the 220$^{\circ}$ - 60$^{\circ}$ longitude sector. Results of the statistical study of the deviation of flux tube crossing times are displayed in Figure \ref{fig8}. 

Figure 9 investigates if the discrepancies between model and data may be an effect of longitude or local time. There is not a clear trend between crossing timing error and the longitude sector in which the moons’ flux tubes crossing happened. However, we observe that the errors of the JRM33+KK2005 estimates are highest for longitudes $\lambda_{\mathrm{S3RH}}$ = 60$^{\circ}$ and $\lambda_{\mathrm{S3RH}}$ = 210$^{\circ}$ while those of JRM33+CON2020 are equally spread with longitudes (Figure 9a).

\begin{figure}[h!]
    \centering
    \includegraphics[width=\textwidth]{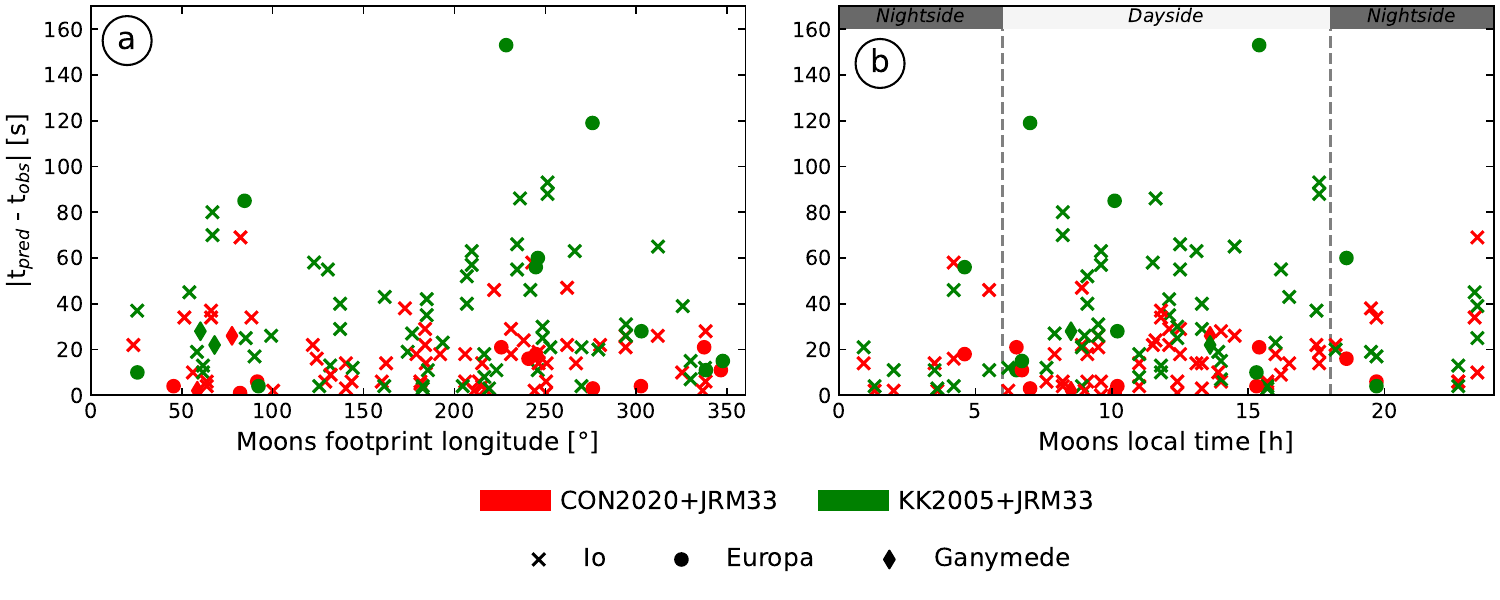}
    \caption{Time delay between observed and predicted flux tube crossings as a function of the longitude sector in which the crossings occurred (a) and the local time of the moon (b). The color code refers to the model used for the estimates, while the markers indicate the moon whose flux tube was crossed by Juno.}
    \label{fig9}
\end{figure}

Conversely, we find a clear correlation between the local time of the moons and the errors of JRM33+KK2005 (Figure \ref{fig9}b). Errors in estimating the crossings of  moons’ flux tubes by Juno are much higher on the dayside (LT = 6h-18h) than on the nightside (LT = 18h-6h). These observations may reveal that the structure of the current sheet is well reproduced by KK2005 on the nightside, especially thanks to the inclusion of  the current sheet deformation in the antisolar direction, but is slightly different than modeled on the dayside.

\section{Concluding remarks and implications for futures studies}

In the present study, we compared the CON2020 and KK2005 current sheet models that reproduce the magnetic field components created by Jupiter’s current sheet combined with the JRM33 intrinsic magnetic field model. Using Galileo measurements acquired over a wide range of radial distances within the current sheet, we have shown that both models are able to successfully predict the reversal of magnetic field directions (Figure \ref{fig1}). In the outer magnetosphere, JRM33+KK2005 is the best at reproducing the amplitude and inversion of the measured magnetic field, likely because it is the only model to include the current sheet bending at large distances from Jupiter. This result is expected since the CON2020 model was derived without considering magnetic field measurements at radial distance R $>$ 30 $\textit{R}_{\mathrm{J}}$. However, as both models are based on data acquired at R $<$ 30 $\textit{R}_{\mathrm{J}}$, there are no apparent constraints for one model to be more accurate than the other in the middle and inner magnetosphere. Consequently, in these regions, we checked the accuracy of the models by contrasting their estimates of the magnetic field components with Juno and Galileo measurements. We find that in the middle magnetosphere, JRM33+KK2005 matches the in situ measurements better than JRM33+CON2020 (Figure 1). Finally,  in the innermost region of Jupiter’s magnetosphere, both models provide a good match with the magnetic field amplitudes measured by Juno between the orbits of Europa and Io. However, we have shown that the time inversion of the magnetic field is better estimated with JRM33+CON2020 (Figure \ref{fig2}). Following these observations, a hybrid current-layer model could be built, considering both CON2020 and KK2005 according to their domain of validity, and using the Juno and Galileo dataset as done by \citeA{wang_2022}. Such a hybrid model, by retaining the precision offered by CON2020 in the inner magnetosphere and by including a current sheet bent by local time effects in the outer magnetosphere as modeled by KK2005, would more accurately predict the magnetic field in all the magnetospheric regions of Jupiter.

In the framework of moon-magnetosphere interactions studies, we estimated the positions of the magnetic footprints of Io, Europa, and Ganymede using JRM33+KK2005 and JRM33+CON2020. We have found that JRM33+CON2020 provides results more in line with the positions of the MAW spots observed by the UVS instrument than JRM33+KK2005 (Figure \ref{fig3}, \ref{fig5} and \ref{fig6}). We investigated the effects of the moon’s local time on the positions of Io’s footprints, showing that the footpaths predicted by JRM33+KK2005 at with different local times do not overlap (Figure \ref{fig4}). However, this effect is not seen in the available Juno UVS observations of the moon footprints. As the magnetic field appears to be unaffected by local time effects at the orbital locations of Io, Europa, and Ganymede, as shown by the accurate estimates of CON2020 model in the inner magnetosphere (Figure 2), a local time effect on the footprints location might not exist. Nevertheless, we note that effects of the moon’s local time on the power emitted by Io’s footprint have been reported by \citeA{hue_2019} using UVS spectrograph data. Further observations of MAW spots on Jupiter’s pole during Juno's next orbits could help to constrain possible moons’ local time effects.    

We used both models to back-trace Juno’s position onto equatorial and magnetic planes during the 12$^{\mathrm{th}}$ perijove of Juno. We have found that Juno’s equatorial projection and M-Shell values are slightly higher than expected with JRM33+KK2005 during the crossings of Europa’s and Io’s flux tubes (Figure \ref{fig7}). 

As a result of the separation between the actual positions of the MAW spots and the footpaths derived by JRM33+KK2005, we showed that the time estimates of moons’ tubes crossings by Juno are closer to the times observed in Juno’s data with the CON2020 current sheet model (Table 1). In Figure \ref{fig8}, we provided a statistical analysis of time differences between estimates and observations of moons’ flux tubes crossings by Juno. 

In the absence of a sufficiently large dataset concerning Callisto’s UV auroral footprints, we did not compare the footpaths derived with both models for this moon. We however point out that Callisto’s radial distance (r$_{\mathrm{Callisto}}$ = 26.3 ± 0.2 $\textit{R}_{\mathrm{J}}$) is close to the limit of validity of the current sheet axisymmetry hypothesis done in the CON2020 model. At such distances, the local time is likely to have an influence on the shape of the current sheet, inducing, for example, its bending. We have shown indeed that in the middle magnetosphere, JRM33+CON2020 is not the best model to reproduce the magnetic field measured by the Galileo spacecraft (Figure \ref{fig1}). Future studies on Callisto’s footprints should therefore also consider using the KK2005 model to track the predicted position of the moon auroral footprints and likely identify their UV signatures. 

The pairing of the CON2020 current sheet model with JRM33 now reaches an accuracy that makes it a powerful tool for predicting and studying the moon-magnetosphere interactions at high latitudes for Io, Europa, and Ganymede. The model could for example be used in order to explore existing databases of UV auroral emissions such as the Auroral Planetary Imaging and Spectroscopy \cite<APIS,>{lamy_2015} where the locations of most of the moon footprints could be predicted, validated, and analyzed. Additional studies constraining the origin of the charged particles source of the most poleward auroral emissions are still needed in order to confirm the validity and accuracy of any current sheet models in the outermost parts of Jupiter's magnetosphere. 

We took into account the short-term variability of the properties of the current sheet by using the minimum and maximum values of the parameters $\mu_{0}I/2$ and $\mu_{0}I_{rad}/2\pi$ of the CON2020 model in our calculations, but found no significant differences with the results derived from the mean model. Furthermore, Supplementary Figure S5 shows the auroral footprints computed with the current parameters derived by \citeA{vogt_2017} using Galileo magnetic field measurements. We find no significant difference between the Galileo-Vogt mapping and the Juno-CON2020 mapping. Therefore, we can conclude that a variation of the physical properties of the current sheet between the Galileo and Juno may not be the reason why JRM33+CON2020 better maps moon auroral footprints than JRM33+KK2005. However, a definite conclusion on the imperfect mapping of moon auroral footprints by JRM33+KK2005 would require data on the position of the moon’s footprints at the time of Galileo with a precision as good as Juno-UVS observation, which unfortunately does not exist. Finally, we note variations over several decades of the magnetospheric fields parameters have been reported by \citeA{momoki_2022}. The impact of such long-term variability on the position of the moon auroral footprints may deserve a dedicated study.

\section*{Open Research}

The JRM33 model used in this study is part of the community codes publicly available \cite<\textcolor{blue}{\url{https://zenodo.org/record/7038966}, }>{wilson_community_codes}. Further information on these codes can be found in Wilson et al.(2023). The CON2020 current sheet model is available at \textcolor{blue}{\url{https://zenodo.org/records/10138957}} \cite{provan_con2020}. The KK2005 model was initially coupled with the VIP4 internal magnetic field model. The original code is available, in IDL language, on the LASP website. We developed our own Python version of the KK2005 model, which we coupled with the JRM33 internal magnetic field model. This version of the model is available at \textcolor{blue}{\url{https://zenodo.org/records/10102742}} \cite{rabia_kk2005_model}. 

The Juno FGM \cite{juno_fgm_data}, Juno JADE \cite{juno_jade_data}, and Galileo MAG \cite{galileo_fgm_data} data used in this study are publicly available on the Planetary Plasma Interactions node of the Planetary Data System website. The Juno and Galileo orbital information can be found in the corresponding spice kernels publicly available from the Navigation and Ancillary Information Facility \cite<NAIF, >{naif}. Part of the analysis has been done using the publicly-available AMDA tool \cite<\textcolor{blue}{\url{http://amda.cdpp.eu}}, >{amda} provided by CDPP.

\acknowledgments

French co-authors acknowledge the support of CNES for the Juno mission. This study has been partially supported through the grant EUR TESS N°ANR-18-EURE-0018 in the framework of the Programme des Investissements d'Avenir. V. H. acknowledges support from the French government under the France 2030 investment plan, as part of the Initiative d’Excellence d’Aix-Marseille Université – A*MIDEX AMX-22-CPJ-04. This study has also been partially supported by the Programme National PNST of CNRS/INSU co-funded by CNES and CEA and by the Programme National PNP of CNRS/INSU co-funded by CNES and CNRS. 


%
%



\bibliography{agusample}

%
%
%
%
%

\end{document}